\documentclass[aps,prb,final,twocolumn,showpacs,groupedaddress]{revtex4}
\usepackage{amssymb,graphicx}

\begin{document}

\title{Can Kinematic Diffraction Distinguish Order from Disorder?}

\author{Michael Baake}
\email[]{mbaake@math.uni-bielefeld.de}
\affiliation{Fakult\"{a}t f\"{u}r Mathematik, Universit\"{a}t Bielefeld,
Postfach 100131, 33501 Bielefeld, Germany}
\author{Uwe Grimm}
\email[]{u.g.grimm@open.ac.uk}
\affiliation{Department of Mathematics and Statistics,
The Open University,
Walton Hall, Milton Keynes MK7 6AA, United Kingdom}

\date{30 October 2008}

\begin{abstract}
Diffraction methods are at the heart of structure determination of
solids.  While Bragg-like scattering (pure point diffraction) is a
characteristic feature of crystals and quasicrystals, it is
not straightforward to interpret continuous diffraction intensities,
which are generally linked to the presence of disorder. However, based on
simple model systems, we demonstrate that it may be impossible to draw
conclusions on the degree of order in the system from its diffraction
image. In particular, we construct a family of one-dimensional binary
systems which cover the entire entropy range but still share the same
purely diffuse diffraction spectrum.
\end{abstract}

\pacs{61.05.cc  
      61.43.-j  
      61.44.-n 
      02.50.-r} 
\maketitle

The inverse problem of reconstructing a structure from its diffraction
pattern is one of the most important challenges in materials
science. Its degree of complexity increases if one goes beyond simple
periodic systems to cover quasicrystals, modulated structures or
complex alloys. In particular, it has been realised that the Bragg
diffraction alone is generally insufficient for a complete
reconstruction \cite{W,NP,GM}.

Currently, an increasing effort is being made to understand and
utilise the continuous part of the diffraction pattern; see
Refs.~\onlinecite{W,NP} for background and Refs.~\onlinecite{d1,d2}
for recent applications. However, even in the idealised situation of a
perfect diffraction experiment with unlimited resolution, the
reconstruction is generally not unique.  Already in 1944, Patterson
\cite{Patt44} discussed \emph{homometric} point sets, which are point
sets whose kinematic diffraction patterns coincide, and provided
explicit examples to illustrate the ambiguity. It was
demonstrated\cite{GM} that it may be possible to lift the ambiguity,
and thus to determine the structure uniquely, if higher-order
correlations are known. While one can argue that, for structures
originating from systems with pure pair-potential interaction (or
allowing a description by effective pair potentials, compare
Ref.~\onlinecite{SC}), higher-order correlations are determined by
the pair correlations \cite{NBISRW}, this is not generally the case,
and in practice measurements of higher-order correlations are
extremely difficult. The role of phase information in stochastic
systems was investigated in Ref.~\onlinecite{WW}.

Here, we want to go one step further, and compare the diffraction of
various point sets, ranging from deterministic to fully stochastic, in
a parametrised way. We characterise the degree of order by the
corresponding (metric) entropy. As we will see below, it is possible
to construct families of point sets which span an entire entropy range
but share the same kinematic diffraction pattern -- proving that
diffraction is insensitive even to the degree of order in this case.

We start by giving a brief introduction to some basic notions of
mathematical diffraction theory, for a one-dimensional (but relevant)
setting with scatterers placed on integer positions. Although this is
a highly idealised situation that ignores displacement effects, its
practical relevance is well known; see Ref.~\onlinecite{W} and
references therein. Then, we discuss the diffraction of two binary
systems (characterised by two scattering strengths) -- a perfectly
ordered structure based on a specific deterministic sequence, and a
completely random structure based on a coin-tossing experiment. It was
observed earlier \cite{HB00} that these rather different systems share
the same diffraction, and are thus homometric. Finally, we introduce a
`Bernoullisation' procedure to couple coin-tossing disorder to a
perfectly ordered structure, thus producing partially ordered systems
of varying entropy. We employ this procedure to explicitly construct a
family of binary systems which are homometric and cover the entire
available entropy range from the perfectly ordered (entropy~$0$) to
the fully stochastic situation (entropy $\log(2)$). Although these
systems may not occur naturally, they can be made synthetically. In
the simplest scenario, a binary structure can be produced by
sequential deposition of layers consisting of two different materials,
which makes it possible to realise any desired sequence; see
Ref.~\onlinecite{AT} for an example. More complicated structures are
also feasible, see for instance Ref.~\onlinecite{St}, and such
artificial materials with designed physical properties will become
increasingly important.

\paragraph{Diffraction of Dirac combs.}

To keep arguments simple, we consider the diffraction of
one-dimensional systems with point-like scatterers located at integer
points $n\in\mathbb{Z}$. The scattering strengths are given by weights
$w_{n}$ for $n\in\mathbb{Z}$, which we assume to be real for
simplicity (the setting can be extended to complex weights). The
corresponding scattering density is modelled by the \emph{Dirac comb}
\[
    \omega = \sum_{n\in\mathbb{Z}} w_{n}\,\delta_{n} ,
\]
where $\delta_{x}$ denotes the normalised point measure (Dirac
$\delta$) on the real line, located at position $x$. Clearly, all
distances between scatterers are integer valued. This implies that the
\emph{autocorrelation} (or Patterson) measure $\gamma$, assuming its
existence for the moment, is again a Dirac comb on $\mathbb{Z}$,
\begin{equation}\label{eq:auto}
    \gamma = \sum_{m\in\mathbb{Z}} \eta(m)\,\delta_{m} ,
\end{equation}
with the coefficients $\eta(m)$ obtained as the limits
\[
   \eta(m) = \lim_{N\to\infty} \frac{1}{2N+1} 
  \sum_{n=-N}^{N} w_{n} \, w_{n+m} .
\]
The scattering intensity $I(k)$ for wave numbers $k\in\mathbb{R}$ is
then determined by the \emph{diffraction} measure $\widehat{\gamma}$,
the Fourier transform of the autocorrelation $\gamma$, compare
Ref.~\onlinecite{Hof} for background. There are several slightly
different versions of the Fourier transform. We prefer to use
\[
    \widehat{\phi}(k)\, = \int_{\mathbb{R}} e^{-2\pi i k x} \,
    \phi(x)\, \mathrm{d} x
\]
for a Schwartz function $\phi$, and its standard extension to tempered
distributions and measures; see Ref.~\onlinecite{S} for details.

For the case of a one-dimensional crystal with $w_{n}=w$ for all
$n\in\mathbb{Z}$, we have $\eta(m)=w^2$ for all $m\in\mathbb{Z}$,
hence $\gamma=w^2\,\delta_{\mathbb{Z}}$, where we use
$\delta^{}_{\mathbb{Z}}$ as shorthand for the sum
$\sum_{n\in\mathbb{Z}}\delta_{n}$. Its Fourier transform is obtained
by Poisson's summation formula \cite{S},
$\widehat{\delta_{\mathbb{Z}}}=\delta_{\mathbb{Z}}$, which gives
$\widehat{\gamma}=w^2\,\delta_{\mathbb{Z}}$. The diffraction image
thus consists entirely of Bragg peaks, located at integer positions
$k$, with equal diffraction intensities $I(k)=w^2$. The diffraction
spectrum in this case is \emph{pure point}, meaning that it consists
of Bragg peaks only. In general, the diffraction measure may comprise
three different contributions,
\[
    \widehat{\gamma} = \widehat{\gamma}_{\mathsf{pp}} + 
                       \widehat{\gamma}_{\mathsf{sc}} +
                       \widehat{\gamma}_{\mathsf{ac}} 
\]
where $\widehat{\gamma}_{\mathsf{pp}}$ is the pure point part,
consisting of a countable sum of $\delta$ peaks. The term
$\widehat{\gamma}_{\mathsf{ac}}$ corresponds to the \emph{absolutely
  continuous} component, which can be described by a locally
integrable (and often continuous) non-negative function
$I_{\mathsf{ac}}(k)$ of the wave vector $k$. The remainder, if there
is any, is called the \emph{singular continuous} component
$\widehat{\gamma}_{\mathsf{sc}}$.  While it vanishes on the complement
of a set $S$ of measure $0$, even within $S$ it never gives weight to
any single point. When such a component is present, $S$ can thus not
be a countable set. Apart from trivial examples of a diffraction
measure that is concentrated on a line in the plane, or similarly on a
manifold of lower dimension, typical examples for this strange
contribution are diffraction intensities which are supported on a
Cantor set or a dense set. A well-known example for the latter
phenomenon is the Thue-Morse chain. For an appropriate choice of its
scattering strengths, it has a purely singular continuous diffraction
measure; see Refs.~\onlinecite{Kaku72,Q,BG08} for derivations and
Refs.~\onlinecite{AT,St} for applications.

If the Dirac comb on $\mathbb{Z}$ is \emph{periodic}, which means that
there is an integer $p>0$ such that $w_{n+p}=w_{n}$ for all
$n\in\mathbb{Z}$, the diffraction measure is pure point, and supported
on the lattice $\mathbb{Z}/p$. It is again periodic, at least with
period $1$, but not necessarily with any smaller period. As an
example, consider the alternating Dirac comb with $w_{n}=(-1)^{n}$. In
this case, $\eta(m)=(-1)^{m}$ for $m\in\mathbb{Z}$, so the
autocorrelation is
$\gamma=\delta_{2\mathbb{Z}}-\delta_{2\mathbb{Z}+1}$.  By Poisson's
summation formula and elementary properties of the Fourier transform
(such as the behaviour under scaling and the convolution theorem), we
have
\[
    \widehat{\delta_{2\mathbb{Z}+1}}
    = \frac{\cos (2\pi k)}{2}\, \delta_{\mathbb{Z}/2} ,
\]
which leads to the diffraction measure
\[
    \widehat{\gamma} = \frac{1-\cos(2\pi k)}{2}\, 
    \delta_{\mathbb{Z}/2} = \delta_{\mathbb{Z}+1/2} .
\]
In this case, the diffraction spectrum is again pure point, and
consists of Brag peaks of unit intensity at positions $n+1/2$,
hence on a subset of $\mathbb{Z}/2$. The fundamental period
of $\widehat{\gamma}$ is nevertheless still $1$.

To obtain absolutely or singular continuous components, in line with
the classification of Ref.~\onlinecite{BM}, we thus have to go beyond
the periodic situation.

\paragraph{Rudin-Shapiro versus Bernoulli.}

\begin{figure*}
\centerline{\includegraphics[width=\textwidth]{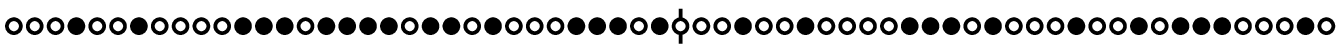}}
\caption{\label{fig:rschain} Central part of the Rudin-Shapiro chain,
with open (full) circles representing scattering strengths $w(n)=1$
($w(n)=-1$), respectively. The location of the origin ($n=0$) is
indicated by the vertical line.}
\end{figure*}

Let us start with a deterministic system without periodicity, based on
the well-known binary Rudin-Shapiro chain. We consider the
corresponding Dirac comb
\[
   \omega^{}_{\mathrm{RS}}=\sum_{n\in\mathbb{Z}}w(n)\,\delta_{n} ,
\]
where $w \! : \, \mathbb{Z} \longrightarrow \{\pm 1\}$ is
defined by the recursion
\begin{equation}\label{eq:rs}
   w(4n+\ell)=
    \cases{ w(n) , & for $\ell\in\{0,1\}$, \cr 
          (-1)^{n+\ell}\,w(n), & for $\ell\in\{2,3\}$,}
\end{equation}
together with the two initial conditions $w(0)=1$ and $w(-1)=-1$. The
resulting system is an aperiodic sequence in $1$ and $-1$, both
appearing equally frequent; see Fig.~\ref{fig:rschain} for a graphical
representation. It has many nice properties, such as strict ergodicity
and linear patch counting complexity, see Ref.~\onlinecite{Q} and
references therein for details. In particular, these properties imply
that this sequence has topological (and metric) entropy $0$.

Since $\omega^{}_{\mathrm{RS}}$ is a Dirac comb on $\mathbb{Z}$, the
autocorrelation is of the form (\ref{eq:auto}), with coefficients
\[
   \eta^{}_{\mathrm{RS}}(m)\,=\lim_{N\to\infty}\frac{1}{2N+1}
   \sum_{n=-N}^{N}w(n)\,w(n+m) .
\]
It follows from the unique ergodicity of the RS sequence that all
these coefficients (and hence $\gamma$) exist, by an application of
the ergodic theorem. They are given by
$\eta^{}_{\mathrm{RS}}(m)=\delta^{}_{m,0}$, which is astonishing.  The
construction of a deterministic sequence with vanishing two-point
correlations was the original (and independent) motivation of Rudin
and Shapiro, thus answering a question in the theory of Fourier
series.

Let us prove this property by a simple, explicit argument. Consider
$a_{m}=\eta^{}_{\mathrm{RS}}(m)$ together with
\[
   b_{m}\,=\lim_{N\to\infty}\frac{1}{2N+1}
   \sum_{n=-N}^{N}(-1)^{n}\, w(n)\,w(n+m) ,
\]
which also exist (by another application of the ergodic theorem).
Clearly, one has $a_{0}=1$ and $b_{0}=0$, because $w(n)^2=1$ for all
$n\in\mathbb{Z}$. Then, considering $m$ modulo $4$, and splitting the
sums in the definition of $a_{m}$ and $b_{m}$ accordingly, the
recursion relations (\ref{eq:rs}) imply 
\begin{eqnarray*}
a_{4m}&=&\textstyle\frac{1+(-1)^m}{2}\,a_{m},\qquad a_{4m+2}\, =\, 0,\\
a_{4m+1} &=& \textstyle\frac{1-(-1)^m}{4}\,a_{m} +
\textstyle\frac{(-1)^m}{4}\,b_{m} - \textstyle\frac{1}{4}\,b_{m+1}\\
a_{4m+3} &=& \textstyle\frac{1+(-1)^m}{4}\,a_{m+1} - 
\textstyle\frac{(-1)^m}{4}\,b_{m} +
\textstyle\frac{1}{4}\, b_{m+1}.
\end{eqnarray*}
Similarly, one finds 
\begin{eqnarray*}
b_{4m}&=&0, \qquad b_{4m+2} = \textstyle
\frac{(-1)^m}{2}\, b_{m}+\frac{1}{2}\, b_{m+1},\\ 
b_{4m+1}&=&\textstyle\frac{1-(-1)^m}{4}\,a_{m} -
\textstyle\frac{(-1)^m}{4}\,b_{m} +\textstyle\frac{1}{4}\,b_{m+1},\\ 
b_{4m+3}&=&-\textstyle\frac{1+(-1)^m}{4}\,a_{m+1} - 
\textstyle\frac{(-1)^m}{4}\,b_{m} +
\textstyle\frac{1}{4}\,b_{m+1}.
\end{eqnarray*}  
Using the initial data, these recursion relations imply that
$a_{m}=b_{m}=0$ for all integers $m\ne 0$. This result shows that the
autocorrelation and diffraction measures of the binary Rudin-Shapiro
Dirac comb $\omega^{}_{\mathrm{RS}}$ are simply
\[
    \gamma^{}_{\mathrm{RS}} = \delta^{}_{0}
    \quad \text{and} \quad 
    \widehat{\gamma^{}_{\mathrm{RS}}} = \lambda ,
\] 
where $\lambda$ denotes Lebesgue measure. In other words, the
diffraction measure is purely absolutely continuous, and consists of a
constant background (of height $1$) only. The extinction of all Bragg
peaks is due to the balanced choice of weights. This is convenient for
the theoretical argument, but also relevant in practice \cite{DF} when
disregarding thermal displacement. Note that all arguments can be
extended to mixed spectra.

Perhaps the most elementary stochastic system is based upon the
classic coin-tossing (or Bernoulli) experiment. We consider a
stochastic Dirac comb\cite{BM98} on $\mathbb{Z}$,
\[
   \omega^{}_{\mathrm{B}}=\sum_{n\in\mathbb{Z}} 
    W_{n}\,\delta^{}_{n} ,
\]
where $(W_{n})^{}_{n\in\mathbb{Z}}$ is a family of independent
identically distributed (i.i.d.) random variables, with probabilities
$\mathbb{P} (W_n = 1) = p$ and $\mathbb{P} (W_n = -1) = 1-p$, where
$0\le p\le 1$.  The corresponding (metric) entropy is
\begin{equation}\label{eq:entropy}
   H(p) =  - p\log (p) - (1-p) \log (1-p),
\end{equation}
which satisfies $0\le H(p)\le \log(2)$. It attains the extremal values
for $p=0$ and $p=1$, where $H=0$ is minimal, and for $p=\frac{1}{2}$,
where $H=\log (2)$ is maximal.

The autocorrelation $\gamma^{}_{\mathrm{B}}=\sum_{n\in\mathbb{Z}}
\eta^{}_{\mathrm{B}}(m)\,\delta^{}_{m}$ is once again a pure point
measure that is supported on $\mathbb{Z}$, with autocorrelation
coefficients given by
\begin{equation}\label{eq:etab}
   \eta^{}_{\mathrm{B}}(m)\, = \lim_{N\to\infty}\,
   \frac{1}{2N+1}\sum_{n=-N}^{N} W_{n}\,  W_{n+m}
\end{equation}
for $m\in\mathbb{Z}$. These coefficients almost surely exist, by an
application of the strong law of large numbers (see below), for all
$m\in\mathbb{Z}$, and satisfy
\[
   \eta^{}_{\mathrm{B}}(m)=\cases{
   1 , & $m = 0$ ,\cr
   (2p-1)^2, & $m\ne 0$ .}
\]
This statement can be proved as follows. It obviously holds for $m=0$,
so consider some fixed $m\ne 0$.  The products $Z_{n}:=W_{n} W_{n+m}$
form a family $\bigl(Z_{n}\bigr)_{n\in\mathbb{Z}}$ of identically
distributed random variables, which take the values $1$ and $-1$ with
probabilities $p^{2}+(1-p)^{2}$ and $2p(1-p)$, respectively.  These
new random variables are not independent, but we can split the sum in
Eq.~(\ref{eq:etab}) into two sums (for instance according to even and
odd values of $[\frac{n}{m}]$, the largest integer smaller than or
equal to $\frac{n}{m}$). The resulting two sums each comprise pairwise
independent random variables. An application of the strong law
of large numbers in its formulation by Etemadi \cite{E} then shows
that each sum almost surely converges (as $N\to\infty$) to the
expectation value of any of the single random variables in the sum,
which is $\frac{1}{2}(2p-1)^2$.  Hence, the diffraction measure of the
stochastic Dirac comb $\omega^{}_{\mathrm{B}}$ almost surely is
\[
   \widehat{\gamma^{}_{\omega^{}_{\mathrm{B}}}\!}\, = 
    \, (2p-1)^{2}\delta^{}_{\mathbb{Z}} + 
   4 p (1-p)\,\lambda ,
\]
where $\lambda$ again denotes Lebesgue measure. In other words, the
diffraction spectrum comprises a constant background of intensity
$4p(1-p)$ for any value of the wave number $k$ and Bragg peaks of
intensity $(2p-1)^2$ at integer $k$. Note that, for the perfectly
ordered cases $p=0$ and $p=1$, the background vanishes, while the
Bragg peaks vanish for the maximally disordered case $p=\frac{1}{2}$.
At this value of $p$, the diffraction measure coincides with that of
the Rudin-Shapiro chain.

This establishes the homometry of the deterministic binary
Rudin-Shapiro chain (with entropy $0$) and the completely random
Bernoulli chain with $p=\frac{1}{2}$ (with entropy $\log(2)$), as
originally observed in Ref.~\onlinecite{HB00}. Coupling the two
systems in a suitable way, we now extend this to an entire
family that covers the intermediate entropy range.

\paragraph{Bernoullisation.}

The Bernoulli chain discussed above is an example of a completely
random and interaction-free system. In view of real world examples,
it is interesting to explore what happens if one imposes the influence
of coin tossing on the order of a deterministic system. This can be
realised in many different ways. Here, we focus on binary sequences
and modify them by an i.i.d.\ family of Bernoulli variables.

Consider a bi-infinite binary sequence $S\in\{\pm 1\}^{\mathbb{Z}}$
which we assume to be uniquely ergodic. Then, the corresponding Dirac
comb $\omega^{}_{S} = \sum_{n\in\mathbb{Z}} S_{n}\,\delta_{n}$
possesses the unique (natural) autocorrelation
$\gamma^{}_{S}=\sum_{m\in\mathbb{Z}}\eta^{}_{S}(m)\,\delta_{m}$ with
the autocorrelation coefficients $\eta^{}_{S}(m)$, where
$\eta^{}_{S}(0)=1$ by construction.

Let $(W^{}_{n})^{}_{n\in\mathbb{Z}}$ be an i.i.d.\ family of random
variables that each take values $+1$ and $-1$ with probabilities $p$
and $1-p$.  The `Bernoullisation' of $\omega^{}_{S}$ is the random
Dirac comb
\begin{equation}\label{eq:bernoullisation}
     \omega\, := \sum_{n\in\mathbb{Z}} S_{n}W_{n}\,\delta_{n},
\end{equation}
which emerges from $\omega^{}_{S}$ by independently changing the sign
of each $S_{n}$ with probability $1-p$. Setting $Z_{n}:=S_{n}W_{n}$
defines a new family of independent (though in general not identically
distributed) random variables, with values in $\{\pm 1\}$. Despite
this modification, the autocorrelation $\gamma$ of $\omega$ almost
surely exists and can be determined via its autocorrelation
coefficients $\eta(m)$ as follows. Since one always has
$\eta(0)=\eta^{}_{S}(0)=1$, let $m\ne 0$ and consider, for large $N$,
the sum
\begin{eqnarray*}
   \lefteqn{ \frac{1}{2N+1} \sum_{n=-N}^{N} Z_{n} Z_{n+m}
    =}\\
    && \frac{1}{2N+1} \left( \sum_{(+,+)} + \sum_{(-,-)} - 
        \sum_{(+,-)} - \sum_{(-,+)}
       \right) W_{n} W_{n+m} ,
\end{eqnarray*} 
which is split according to the value of $(S_{n},S_{n+m})$. Each of
the four sums can then be handled in the same way as in the argument
for the Bernoulli chain above, thus contributing $(2p-1)^2$ times the
frequency of the corresponding sign pair. Observing that the overall
signs are the products $S_{n}S_{n+m}$, it is clear that, as
$N\to\infty$, one (almost surely) obtains
\[
    \eta(m) = (2p-1)^{2}\,\eta^{}_{S}(m)
\]
for all $m\ne 0$. This shows that the new autocorrelation almost
surely is
\[
   \gamma = (2p-1)^{2}\,\gamma^{}_{S} + 
   4 p (1-p)\, \delta^{}_{0}
\]
where $\gamma^{}_{S}$ is the unique autocorrelation of
$\omega^{}_{S}$.

Let us apply this Bernoullisation procedure to the Rudin-Shapiro
chain. Denote by $\omega$ the random Dirac comb obtained from the
Bernoullisation (with parameter $p$) of the binary Rudin-Shapiro
chain. Then, the autocorrelation measure almost surely exists and
reads $\gamma=\delta^{}_{0}$, \emph{independently} of $p$. This means
that the random Dirac combs $\omega$, even for different values of
$p$, are almost surely homometric, and share the purely absolutely
continuous diffraction measure $\widehat{\gamma}=\lambda$.

Note that this example explores the full entropy range of
Eq.~(\ref{eq:entropy}): the Bernoulli case (with $p=\frac{1}{2}$) has
entropy $\log (2)$, the maximal value for a binary system, while
Rudin-Shapiro has entropy $0$, and the parameter $p$ interpolates
continuously between the two limiting cases.  The solution of the
corresponding inverse problem is thus highly degenerate. Unless
additional information is available, for instance via higher order
correlations, one possible strategy could employ a maximum entropy
method \cite{GS}, singling out the Bernoulli comb.

Both the Bernoullisation procedure and the specific one-dimensional
examples immediately generalise to higher dimensions by taking direct
product structures. In particular, the product of $d$ Rudin-Shapiro
chains results in a deterministic system in $d$-space with the same
purely absolutely continuous diffraction measure as the corresponding
coin-tossing model. Consequently, our above conclusions extend to this
case.  This means that one can also produce examples with lower rank
entropy, which is a new phenomenon in dimensions $d\ge 2$.

\paragraph{Concluding remarks.}

Diffraction methods provide the most important approach to structure
determination. The presence of Bragg diffraction clearly indicates an
ordered structure, though the discovery of quasicrystals \cite{SBGC}
in the 1980s has shown that pure point diffraction occurs in more
general systems than just conventional crystals. To date, the precise
atomic structure of quasicrystalline alloys is still not completely
understood; there is evidence that entropy plays an important role in
stabilising quasicrystalline structures, and that some disorder may be
an inherent feature of these alloys. Like for many ordinary crystals,
diffuse scattering is present in experimental diffraction patterns of
even the best known quasicrystals, and there is an increasing effort
to explore the information contained the diffuse diffraction intensity
\cite{W,NP}; see Ref.~\onlinecite{d1} for a recent example. It is
tempting to draw conclusions about the degree of order in a structure
on the basis of the observed diffuse scattering intensity. However, as
our explicit example demonstrates, such conclusions have to be
considered carefully, since the relation between diffuse scattering
and disorder is far from simple.

\begin{acknowledgments}
This work was supported by the German Research Council (DFG), within
the CRC 701, and by EPSRC via Grant EP/D058465. It is a pleasure to
thank the School of Mathematics and Physics at the University of
Tasmania for their kind hospitality.
\end{acknowledgments}

\end{document}